\documentclass[preprint, authoryear]{elsarticle}

\usepackage{amssymb}
\usepackage{amsmath}
\usepackage{amsthm}
\usepackage{mathrsfs}
\usepackage{graphicx}

\usepackage[colorlinks,citecolor=blue,urlcolor=blue]{hyperref}

\renewcommand{\mathcal}{\mathscr}

\newcommand{\diag}{{\text{diag}}}

\newcommand{\PP}{\mathbb{P}}
\newcommand{\EE}{{\mathbb{E}}}

\newcommand{\RR}{\mathbb{R}}
\newcommand{\NN}{\mathbb{N}}

\newcommand{\ZZ}{\mathbb{Z}}

\newcommand{\qv}[1]{\left<  #1  \right>}

\renewcommand{\phi}{\varphi}

\newcommand{\given}{\,|\,}

\newcommand{\bbP}{\PP}

\theoremstyle{definition}

\journal{Mathematical Biosciences}

\begin{document}

\begin{frontmatter}



\title{Nonparametric Bayesian methods for one-dimensional diffusion models}


\author{Harry van Zanten}

\address{Korteweg-de Vries Institute for Mathematics\\
University of Amsterdam\\
{\tt hvzanten@uva.nl}}

\begin{abstract}
In this paper we review recently developed methods for 
nonparametric Bayesian inference for one-dimensional diffusion models. 
We discuss different possible prior distributions, computational issues, 
and asymptotic results. 
\end{abstract}


\end{frontmatter}

\numberwithin{equation}{section}

\section{Introduction}

Stochastic Langevin models or diffusion models arise in many fields of applied science. 
Basically they describe the evolution of a system whose 
dynamics are governed by a ``noisy'' ordinary differential equation. 
If $X_t \in \RR^d$ denotes the state of the system at time $t$, then the general form 
of such an equation is 
\begin{equation}\label{eq: presde}
\frac{dX_t}{dt} = b(t,X_t) + \text{``random noise''}.
\end{equation}
The function $b$ describes the instantaneous drift of the process $X$. This 
drift is perturbed by random noise, with an intensity that can be time and state dependent in general. 
A more detailed, 
but still informal description and many examples of applications in the context 
of molecular modeling can for instance be found in Section 14.4 of \cite{Schlick}. 

As explained in Section 14.5 of \cite{Schlick}, 
we can view (\ref{eq: presde}) as an informal description of the 
stochastic differential equation (SDE)
\begin{equation}\label{eq: sde1}
dX_t = b(t,X_t)\,dt + \sigma(t, X_t)\,dW_t.
\end{equation}
Here $W$ is a Brownian motion, which models the random noise,  and the diffusion function $\sigma$ describes
the impact of the current time and state on the level of the noise (see the next section for more details). 
In this paper we restrict our attention to one-dimensional models, i.e.\ models for which 
the state $X_t$ is real-valued. Moreover, we consider only  time-homogenous 
SDEs, meaning that $b$ and $\sigma$ are only functions of the state. 
Such one-dimensional diffusion models are applied in many different fields in the biosciences.
Often they arise after a  reduction of a higher-dimensional system to dimension one, 
achieved for instance by suitable aggregation of data, 
or by a principle component analysis. 
Some concrete examples include models for the membrane potential in a neuron (e.g.\ \cite{Lansky}), 
population size models (\cite{Fleming}), decision making models (\cite{Roxin}), 
reduced models for neurodynamical data (\cite{Deco}) and angle models in molecular dynamics (\cite{Papa}). 

Fitting a diffusion model to observed data amounts to estimating the  functions $b$ and $\sigma$. 
In certain cases it is reasonable to postulate a specific parametric form of these functions, i.e.\ to assume that 
they are known up to some finite-dimensional parameter. 
A well-known example is the mean-reverting Ornstein-Uhlenbeck process, for which 
$\sigma$ is constant and $b(x) = \alpha(x-\beta)$ for some $\alpha < 0$ and $\beta \in \RR$. 
This is in fact the classical Langevin equation, cf.\ \cite{Langevin}. 
Fitting this model reduces to estimating the parameter  $\theta = (\alpha, \beta, \sigma) \in \RR_+ \times \RR \times \RR_+$. 
See for instance \cite{Kut} and \cite{Kessler} for an overview of parametric methods for SDEs. 

In certain cases however, natural parametric specifications are not possible or undesirable and one has to resort to nonparametric methods for making inference 
on the functions $b$ and  $\sigma$.
Several such methods have been proposed in the literature. 
An incomplete list include kernel methods 
(e.g.\ \cite{Banon}, \cite{Kut}), 
penalized likelihood methods (e.g.\ \cite{Comte}), and spectral approaches  \cite{Bandi}.

The statistical techniques mentioned thus far are all ``frequentist'' in nature, i.e.\ non-Bayesian.
In almost all branches of applied statistics however, the use of Bayesian  methods has hugely increased in 
recent years. This is to a large extent due to the development of computational methods. 
Another appeal is that a Bayes procedure provides 
a natural way to quantify the uncertainty in the estimates, through the spread of the posterior distribution. 
Moreover,  the construction of a prior distribution, as required in the Bayesian paradigm,  can be a 
useful tool to bring structure to a complex statistical model. 
Initially the use of Bayes methods in nonparametric problems was met with skepticism, since it 
was pointed out early on that prior specification is a delicate matter in these cases (e.g.\ \cite{Free63, Free65}).
However, mathematical and practical insights of the last decade have shown that these difficulties 
can be overcome. As a result Bayesian methods are now widely used in problems like nonparametric 
regression, density estimation and classification, in many different application areas (see for instance \cite{BNP} for an overview). 

The development and study of Bayesian methodology for SDEs started relatively late 
and initially focussed on parametric models. 
See for instance the papers \cite{Eraker}, \cite{RobertsStramer}, \cite{BeskosPapaspiliopoulosRoberts},
 to mention but a few. 
Only a handful of papers about 
nonparametric Bayes methods for SDEs are available at  the present time. The first paper to  
 propose a practical method  is \cite{Papa}. The theoretical, asymptotic behavior of the procedure
of \cite{Papa} is studied in the  paper \cite{PokernStuartvanZanten}. 
\cite{Moritz} recently proposed an alternative computational approach. 
Other available papers deal with asymptotics in this framework, cf.\  \cite{MeulenVaartZanten}, 
\cite{PanzarVZanten} and \cite{VdMeulenVZanten}, \cite{Shota}.

In the present paper we review the recently developed nonparametric Bayes methods  for scalar SDEs. 
An interesting aspect of these methods is that they provide a natural way
for dealing properly with low-frequency data. Moreover, they allow to report 
credible bands for uncertainty quantification in addition to an estimator of the function of interest. 
The few examples that exists at the moment show that the methods have become numerically feasible. 
As a result, nonparametric Bayes methods can be expected to 
become more and more common tools for fitting SDE models to observed data. 

Throughout the paper a balance is sought between mathematical rigor and a lucid presentation. 
This means that statements are sometimes loosely formulated, or that regularity conditions are taken 
for granted. We give references to the literature for readers seeking more mathematical detail.

The remainder of the paper is organized as follows. 
In the next section we briefly treat some facts from the theory of stochastic differential equations, mainly to 
recall some terminology and fix notation. 
In Section \ref{sec: bayes} we discuss generalities about doing nonparametric Bayesian inference for diffusions. 
In particular, the differences between continuous and low-frequency data are outlined. 
Recently proposed concrete priors are considered in Section \ref{sec: priors}. Section \ref{sec: asymp}
gives an overview of the available asymptotic theory. We end with some concluding remarks in Section \ref{sec: con}.

\section{One-dimensional SDEs}

In this section we very briefly review some relevant theory of one-dimensional stochastic differential equations. 

\subsection{Brownian motion}

The fundamental building block for stochastic differential equations is the Brownian motion process. 
Formally, a collection of random variables  $W = (W_t: t \ge 0)$ defined on a common probability space 
is called a (standard) Brownian motion if 
\begin{enumerate}
\item
$W_0 = 0$, 
\item
for all $t \ge s \ge 0$, $W_t- W_s$ is independent of $(W_u: u \le s)$,
\item
for all $t \ge s \ge 0$, $W_t- W_s$ has a normal  distribution with mean $0$ and variance $t-s$,  
\item
almost every sample path $t \mapsto W_t$ is continuous. 
\end{enumerate}
Brownian motion plays a crucial role in stochastic process theory and has been and still is studied extensively. 
See for instance the books \cite{Revuz} and \cite{Peres} and the many references therein. 

It follows immediately from the definition that the Brownian motion $W$ is 
a Gaussian process with mean $\EE W_t = 0$ and covariance $\EE W_s W_t = \min\{s, t\}$ 
for all $s, t \ge 0$. 
Although it can proved that the ordinary derivative of Brownian does not exist, it can be thought of 
as the primitive of Gaussian white noise. As such it is instrumental in putting 
the loosely described ``ODE with noise'' \eqref{eq: presde} on a firm mathematical basis.

\subsection{Stochastic differential equations}

Mathematically, the stochastic differential equation \eqref{eq: sde1} is shorthand notation for the 
corresponding integral equation
\begin{equation}\label{eq: sie}
X_t  = X_0 + \int_0^t b(s, X_s)\,ds + \int_0^t \sigma(s, X_s)\,dW_t. 
\end{equation}
Here  the second integral is a  stochastic integral, which 
has to be carefully defined and which
obeys  different calculus rules than
ordinary integrals do. 
In particular, the main rule of calculus is replaced by It\^o's formula, which states that if $X$ 
solves \eqref{eq: sde1} and $f$ is a twice continuously differentiable  function, then $f(X)$ 
satisfies the SDE
\[
df(X_t) = (f'(X_t)b(t, X_t) + \frac12 f''(X_t)\sigma^2(t, X_t))\,dt + f'(X_t)\sigma(t, X_t)\,dW_t.
\]
See any  text on stochastic calculus for details (e.g.\ \cite{Chung}, 
\cite{Karatzas}, \cite{Oksendal}).
The stochastic integral in \eqref{eq: sie} is well approximated (in probability) by the so-called 
Riemann-It\^o sum 
\[
\sum_{i=1}^n \sigma((i-1)t/n, X_{(i-1)t/n})(W_{it/n}-W_{(i-1)t/n})
\]
if $n$ is large enough.

It can be proved that under certain regularity conditions on the functions $b$ and $\sigma$, 
for instance the classical Lipschitz and linear growth conditions, there exists a unique stochastic 
process $X$ which solves the integral equation \eqref{eq: sie} (e.g.\ \cite{Karatzas}).
Moreover, this solution is adapted to the Brownian motion $W$, in the sense that 
for every $t \ge 0$, $X_t$ only depends on $(W_s: s \le t)$, i.e.\ it does not ``look into the future''.

The notation  \eqref{eq: sde1} is reasonable since  it correctly describes the 
infinitesimal behavior of the solution $X$ of \eqref{eq: sie} in the sense that loosely speaking we have for very small $h > 0$ that 
\[
X_{t+h} \approx X_t + b(t, X_t)h + \sigma(t, X_t)(W_{t+h}-W_t). 
\]
By the properties of the Brownian motion and the fact that $X$ is adapted, we thus have that 
in distribution, 
\[
X_{t+h} \approx X_t + b(t, X_t)h + \sqrt{h}\sigma(t, X_t)Z,  
\]
where $Z$ is a standard normal random variable independent of $X_t$. This gives a basic method
for recursively simulating solutions to  SDEs, the so-called Euler scheme.
See for instance \cite{Kloeden} for (much) more on this topic.

\subsection{Girsanov's theorem}
\label{sec: girsanov}

Below we will restrict our attention to the solution of a stochastic differential equation 
of the  form 
\[
dX_t = b(X_t)\,dt + dW_t, \qquad X_0 = x_0. 
\]
For $T > 0$, the  restriction $X = (X_t: t \in [0,T])$ of this process to the interval $[0,T]$
is a random element of the space $C[0,T]$ of continuous functions on $[0,T]$. 
As such it generates a distribution, or law, $\PP^T_b$ on this function space, defined by 
$\PP^T_b(B) = \PP(X \in B)$ for (Borel) subsets of $C[0,T]$. 

Girsanov's theorem implies that the distribution $\PP^T_b$ has a density relative to the distribution $\PP_0^T$
of the Brownian motion $W=(W_t: t \in [0,T])$. Moreover, we have an expression for the corresponding 
Radon-Nikodym derivative, or, in statistical terminology, the likelihood. We have
\[
\frac{d\bbP^T_b}{d\bbP^T_0}(X)=\exp\Bigl(- \frac12 \int_0^T b^2(X_t)\, dt + \int_0^T  b\bigl(X_t\bigr)\,dX_t\Bigr)
\]
(see for instance  \cite{LiptserShiryayevI}).

\section{Bayesian inference for SDEs}
\label{sec: bayes}

Suppose we observe a stochastic process $X$ which solves a one-dimensional SDE 
of the form 
\begin{align*} 
dX_t = b(X_t)\,dt + \sigma(X_t)\,dW_t, 
\end{align*}
where $W$ is a Brownian motion and $b$ and $\sigma$ are functions satisfying 
certain regularity conditions ensuring at least that the SDE has a unique (weak) solution for 
every initial condition. Say we observe the process up till some time $T > 0$.
 
Estimating the diffusion function $\sigma^2$ is a degenerate statistical problem, at least if the data are recorded continuously, 
in the sense that its restriction to the range of the data can be recovered 
without error. We can use  the fact that the quadratic variation of the process $X$ is given by 
\[
\qv{X}_t = \int_0^t \sigma^2(X_s)\,ds. 
\]
The left-hand side of this equation is a measurable function of the data, we have for instance that 
as $n \to \infty$, 
\[
\sum_{i=1}^n (X_{it/n} - X_{(i-1)t/n})^2 \to \qv{X}_t
\]
in probability, for every $t \ge 0$ (e.g.\ \cite{Jacod}).
Assuming therefore that $\sigma$ is known we can 
instead of the original process $X$ consider the transformed process 
\[
\int_0^{X_t} \frac1{\sigma(x)}\,dx, \quad t \ge 0.
\]
By It\^o's formula this process has unit diffusion and the statistical problem reduces to making 
inference about its drift function. 

In view of these observations we will assume throughout the remainder of the paper that 
the SDE under consideration has unit diffusion and focus on estimating the drift. 
In the case of low-frequency data,  the transformation outlined above can not be carried out however, and the  
extension of the methods we discuss ahead to the case of an unknown
diffusion function is not always straightforward. 
We refer to \cite{RobertsStramer} for an approach that allows a parametric description of the diffusion function.

\subsection{Continuous-time observations}

Suppose that for $T > 0$, we observe the unique solution $X = (X_t: t \in [0,T])$ of the SDE 
\begin{equation}\label{eq: sde}
dX_t  = b(X_t)\,dt + dW_t, \qquad X_0 = x_0, 
\end{equation}
where $W$ is a Brownian motion and $b: \RR \to \RR$ is an unknown, continuous drift function.  
By Girsanov's theorem, the law that the process $X$ generates on the space $C[0,T]$ of 
continuous functions on $[0,T]$ is equivalent to the Wiener measure on the space (with the appropriate initial condition) and  
the corresponding likelihood  satisfies
\begin{equation}\label{eq: lik}
p(X \given b) = \exp\Bigl(- \frac12 \int_0^T b^2(X_t)\, dt + \int_0^T  b\bigl(X_t\bigr)\,dX_t\Bigr)
\end{equation}
(see Section \ref{sec: girsanov}). 

The nonparametric Bayesian approach now consists in  putting  a prior distribution $\Pi$ on the ``parameter'' $b$
and computing the corresponding posterior distribution. 
Formally  the prior $\Pi$ can be any probability measure on the space $C(\RR)$ of continuous functions on $\RR$
or on a suitable subspace, for instance a space of functions with a certain regularity, and/or certain periodicities. 
The posterior distribution of $b$, which we denote by  $\Pi(\cdot \given X)$, is then given by 
the usual Bayes formula
\[
\Pi(b \in B\given X) = \frac{\int_B p(b\given X)\Pi(db)}{\int p(b\given X)\Pi(db)}.
\]
(In concrete situations it has to be verified that the integrands are properly measurable, so that the integrals 
are well defined.)

The integrals in the expression for the posterior are over infinite-dimensional spaces, which makes
it challenging to do computations. We will see in Section \ref{sec: priors} ahead however that various sensible choices 
for the prior allow the construction of feasible algorithms for drawing realizations from the posterior.

In general the drift  $b$ in (\ref{eq: sde}) is a function defined on the whole real line. This makes 
it not completely obvious to come up with  reasonable priors, as most priors available in the literature 
are defined on spaces of compactly supported functions. 
However, we can usually  work with such more common priors in the SDE case as well. 
In certain applications it is natural to assume that $b$ is a periodic function, reducing the problem 
to estimating a function on $[0,2\pi]$, or another finite interval.  This is for instance the case if 
the available data consist of recordings of angles (e.g. \cite{Yvo}, \cite{Hindriks} or \cite{Papa}). 
But also if periodicity can not be assumed it is typically 
 only sensible to estimate the  function $b$ on a compact interval $I \subset \RR$, 
 since far away from the range of the data there is simply no information available about the function. 
Now note that if we define, for a set $S \subset \RR$,
\[
b_S(x) = \begin{cases}
b(x) & \text{if $x \in S$},\\
0 & \text{else},
\end{cases}
\]
then, with $I^c$ denoting the complement of the interval $I$, the likelihood factorizes as  
$p(X \given b)  = p(X\given b_I)p(X\given b_{I^c})$. It follows that if we put a prior on $b$ 
by putting independent priors $\Pi_I$ and $\Pi_{I^c}$ on $b_I$ and $b_{I^c}$, respectively, 
then the marginal posterior for $b_I$ does not depend on the prior $\Pi_{I^c}$ and is given by 
\[
\Pi(b_I \in B\given X) = \frac{\int_B p(b_I\given X)\Pi_I(db)}{\int p(b_I\given X)\Pi_I(db)}.
\]
Hence in this case as well, we only need to specify a prior on a compactly supported function. 

In the examples in Section \ref{sec: priors} we shall consider the periodic case, 
but simple modifications allow to deal with the non-periodic case as well.

\subsection{Low-frequency observations}

In the preceding subsection  we have been dealing with continuously observed diffusions. 
Obviously, the phrase ``continuous data'' should be interpreted properly. 
In practice it means that the frequency at which the diffusion is observed is so high that 
the error that is incurred by approximating the  stochastic and ordinary integrals like the 
ones appearing in the likelihood \eqref{eq: lik} 
by the corresponding Riemann or Riemann-It\^o sums, is negligible. 
If we only have low-frequency, discrete-time observations at our disposal, these approximation errors can 
typically not be ignored however and can  introduce undesired biases.

Assume now that we only have partial observations 
$X_0, X_\Delta, \dots, X_{n\Delta}$ of the solution of \eqref{eq: sde}, for some $\Delta > 0$ and $n \in \NN$.  
We set $T = n\Delta$. 
Under mild regularity conditions the discrete observations constitute a Markov chain, but it is well known that 
the transition densities of discretely observed diffusions 
and hence the likelihood is not available in closed form in general. This complicates a Bayesian analysis.
An approach that has been proven to be 
very fruitful is to view the continuous diffusion segments 
between the observations as missing data and to treat them as latent (function-valued) variables. 
Since the continuous-data likelihood is known (cf.\ the preceding subsection), 
data augmentation methods (see \cite{TannerWong}) can be used to circumvent the unavailability of the likelihood 
in this manner. 

Concretely, let us again denote the full, continuous-time process up till time $T$ by 
$X = (X_t: t \in [0,T])$. Asume that we have solved the inference problem for the 
continuous data problem described in the preceding subsection, in the sense that 
we have an algorithm that generates (approximate) draws from the posterior distribution 
$\Pi(\cdot \given b)$ of $b$ given the full data $X$. 
The data augmentation method relies on the fact that in the present situation it is also possible 
to generate draws from  the conditional distribution 
\[
X \given b, X_0, X_\Delta, \ldots, X_{n\Delta}
\]
 of the full process $X$ given the discrete-time observations $X_0, X_\Delta, \ldots, X_{n\Delta}$ 
 (details ahead). 
Approximate draws from the target posterior distribution, i.e. the distribution  of the drift given 
$X_0, X_\Delta, \ldots, X_{n\Delta}$, can then 
be obtained from a Gibbs sampler which is initialized at some function $b$ and then repeats the steps
\begin{enumerate}
\item
draw $X \given b, X_0, X_\Delta, \ldots, X_{n\Delta}$,
\item
draw $b \given X$ 
\end{enumerate}
a large number of times. 

By the Markov property of the diffusion, step 1. in the Gibbs sampler can be 
done by independently drawing the $n$ missing segments 
\begin{equation}\label{eq: bridge}
(X_t: t \in ((i-1)\Delta, i\Delta)) \given b, X_{(i-1)\Delta}, X_{i\Delta}
\end{equation}
for $i = 1, \ldots, n$, and gluing them together to obtain the full path $X$. 
The crucial observation is that the diffusion bridge law \eqref{eq: bridge} is equivalent to 
a Brownian bridge that starts in $X_{(i-1)\Delta}$ at time $(i-1)\Delta$ and ends up in 
$X_{i\Delta}$ at time $i\Delta$. 
Moreover, by Girsanov's theorem again, the 
corresponding Radon-Nikodym derivative is proportional to  
\begin{equation*}
	\exp\Big( \int_{(i-1)\Delta}^{i\Delta} b(X_t)\,dX_t- \frac12 \int_{(i-1)\Delta}^{i\Delta} b^2(X_t)\,d t\Big).
\end{equation*}
Since it is straightforward to simulate Brownian bridges, this makes it possible 
to simulate diffusion bridges using for instance rejection sampling or Metropolis-Hastings techniques.
Exact simulation methods for diffusion bridges have been proposed in the literature as well, see for instance \cite{BeskosPapaspiliopoulosRoberts}, 
\cite{BeskosPapaspiliopoulosRobertsFearnhead}. 
For the present purposes exact simulation is not strictly necessary  however and it is 
usually more convenient to add a Metropolis--Hastings (MH) step corresponding to a Markov chain that has the diffusion bridge 
law given by (\ref{eq: bridge}) as stationary distribution. 
For more details on this type of MH samplers for diffusion bridges 
we refer to  \cite{RobertsStramer}.

\section{Gaussian and conditionally Gaussian priors}
\label{sec: priors}

For successful Bayesian inference for SDEs it is obviously important that a prior is used 
that makes the procedure computationally feasible. Moreover, 
to avoid inconsistency problems, we should aim at using 
priors with ``large support'', in the sense that they do not exclude too many 
drift functions. In this section we review recently proposed options.

\subsection{Finite-dimensional priors}

A first, perhaps naive approach to constructing a prior on the drift function $b$ is to 
choose a finite set of basis functions $\psi_1, \ldots, \psi_m$, 
assume that the drift admits an expansion $b = \sum_{j=1}^m c_j \psi_j$
 and to put a prior distribution on the vector of coefficients $c = (c_1, \ldots, c_m)$. 
In terms of $c$ the likelihood can then be written as
\[
p\Big(\sum c_j\psi_j \given X\Big) = e^{c^T\mu -\frac12 c^T\Sigma c}, 
\]
where the data enters through the vector $\mu$ and matrix $\Sigma$ with components
\[
\mu_j = \int_0^T \psi_j(X_t)\,dX_t, \qquad 
\Sigma_{ij} = \int_0^T\psi_i(X_t)\psi_j(X_t)\,dt,
\]
for $i,j = 1, \ldots, m$. Hence if the prior on the coefficients $c$ has a Lebesgue density $\pi$, then 
the posterior distribution of  $c$ has a density proportional to 
\[
c \mapsto \pi(c) e^{c^T\mu -\frac12 c^T\Sigma c}.
\]
Given draws from this posterior for $c$ we obtain draws from the posterior for $b$ by 
combining the coefficients with the basis functions. 

Since the likelihood is log-quadratic, it is  convenient to choose a Gaussian prior on $c$. It is straightforward
to verify that if the prior on $c$ is $N_m(0, \Lambda)$, i.e.\ an $m$-dimensional 
normal distribution with mean $0$ and covariance matrix $\Lambda$, then the posterior for $c$ is Gaussian 
as well, namely
$N_m((\Sigma +\Lambda^{-1})^{-1}\mu, \quad (\Sigma +\Lambda^{-1})^{-1})$.
Sampling from this posterior distribution is straightforward in principle, although the necessary matrix 
inversions can become numerically demanding as $m$ gets large. 
It can be advantageous to employ basis functions leading to a sparse matrix $\Sigma$, in order 
to speed up the matrix computations.

\begin{figure}[h]
\centerline{\includegraphics[scale=.5]{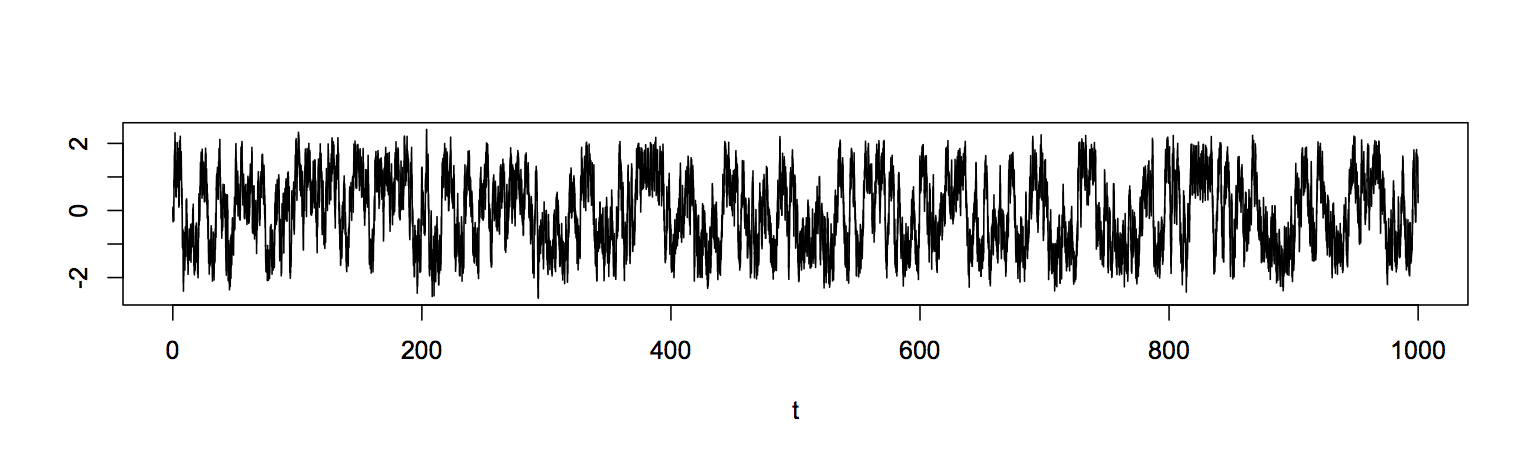}}
\centerline{\includegraphics[scale=.4]{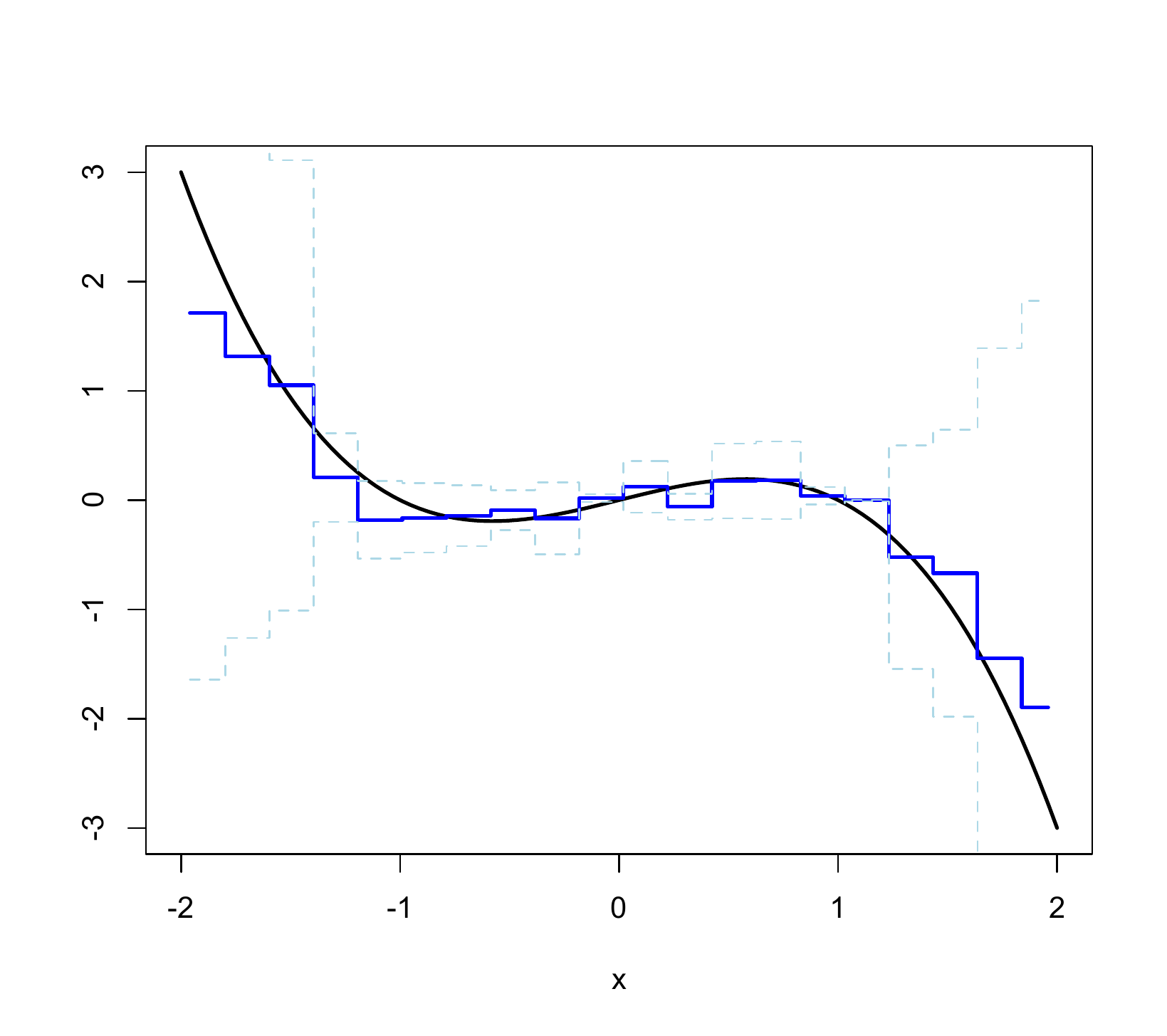}\includegraphics[scale=.4]{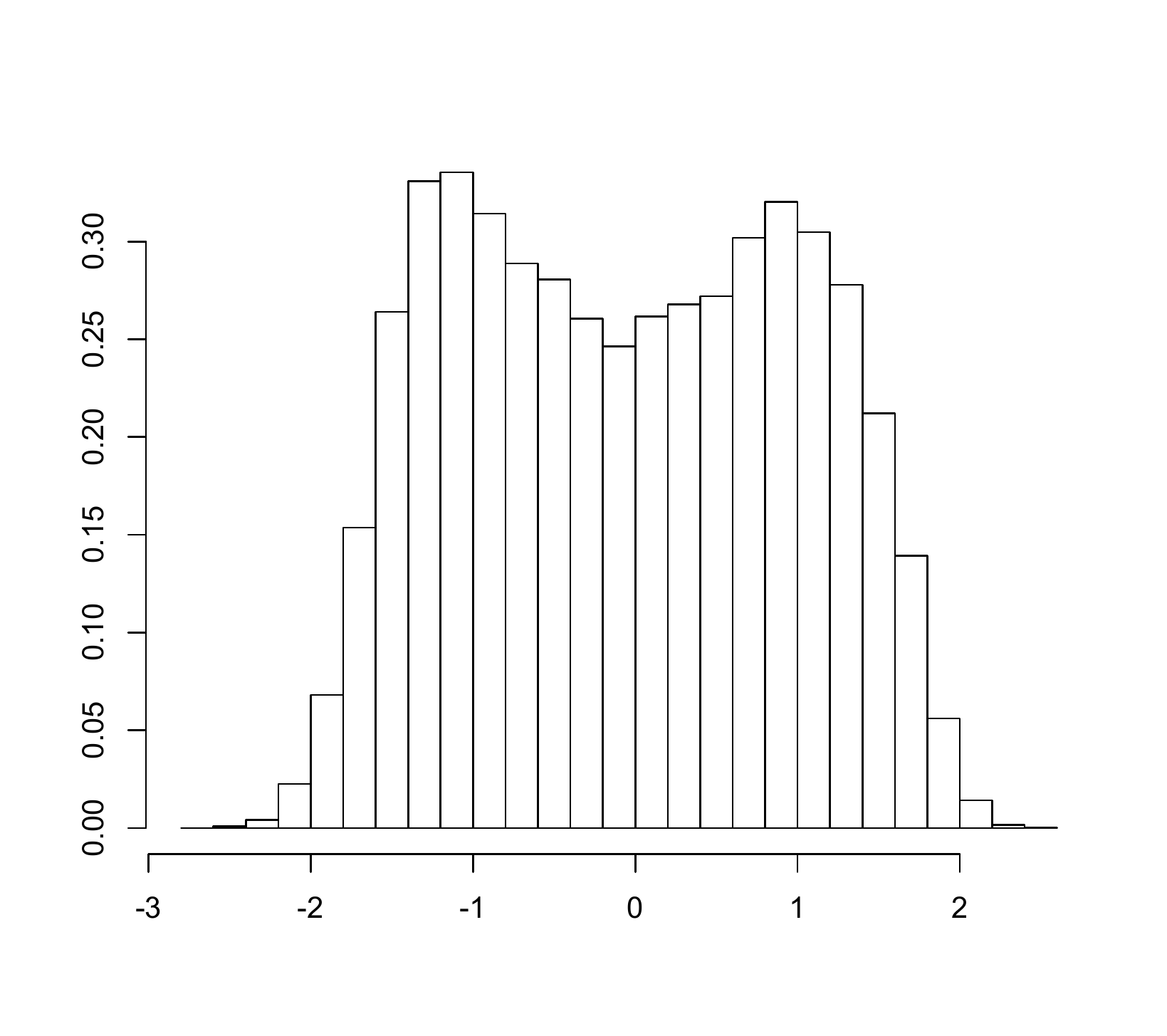}}
\caption{Top panel: simulated data. Lower left: true drift (black), posterior mean (solid blue) and $.95$ pointwise 
credible intervals (dashed blue). Lower right: histogram of the data.}
\label{fig: 1}
\end{figure}

The sketched procedure can work quite well, but only if the drift is in actual fact (close to) a linear 
combination of the chosen basis functions. 
When using a prior on a space of functions with a fixed, finite dimension, only the projection 
of the true drift on this space can be recovered. This can look quite different than the actual drift. 
Figure \ref{fig: 1} illustrates this point. Here we simulated data from the SDE \eqref{eq: sde} with 
drift function $b(x) = -(1/2)x(x-1)(x+1)$. We defined a prior on $b$ by dividing the interval $[-2, 2]$
into $20$ subintervals of equal length and writing $b$ as a linear combination of indicator functions
of these intervals, with independent Gaussian coefficients. The lower left-hand panel of Figure \ref{fig: 1}
visualizes the corresponding posterior. The solid blue line is the posterior mean and the dashed lines 
describe $.95$ point wise credible intervals. Clearly, this posterior only gives a very crude picture
of the true drift (which is the solid black curve). We note that the credible bands are very wide near $-2$ and $2$, 
since only very limited data fall into that region, cf.\ the histogram of the data in the lower right-hand panel of
the figure.

\begin{figure}[h]
\centerline{Figure \ref{fig: 1} about here.}
\end{figure}

%

In the next two subsections we describe two recently proposed procedures 
that avoid this problem by employing truly infinite-dimensional prior distributions for the drift, 
both with large support.

\subsection{Gaussian priors with differential precision operators}
\label{sec: papa}

Let us assume that the drift function $b$ in \eqref{eq: sde} is continuously differentiable, $1$-periodic and 
zero-mean, in the sense that $\int_0^1b(x)\,dx = 0$. For this situation \cite{Papa} propose a 
centered Gaussian prior 
on the space $L^2[0,1]$ of square integrable functions on the unit interval. In general, a centered Gaussian measure
$\Pi$ on $L^2[0,1]$ is determined by its covariance operator $\Lambda : L^2[0,1]
\to L^2[0,1]$ which has the property that 
\[
\int \Big(\int_0^1 g(x)f(x)\,dx\Big)\Big(\int_0^1 h(x)f(x)\,dx\Big)\,\Pi(df) = \int_0^1 g(x)(\Lambda f)(x)\,dx
\]
for all $g,h \in L^2[0,1]$. A linear operator on $L^2[0,1]$ is a covariance operator of a Gaussian measure 
if and only if it is positive definite,  symmetric, and trace-class (e.g.\ \cite{Bogachev}). 
The covariance operator  of \cite{Papa} is defined through its inverse, the so-called precision operator. 
Given fixed hyperparameters $\eta, \kappa > 0$ and $p \in \{2, 3, \ldots\}$ this inverse is the densely defined operator 
\begin{align}
 \Lambda^{-1} = \eta\left( \left(-\Delta\right)^p + \kappa I \right) \label{eq:priorCov},
\end{align}
where $\Delta$ is the one-dimensional
Laplacian, i.e.\ $\Delta f = f''$ and $I$ is the identity operator. The domain of $\Lambda^{-1}$ is 
 the  space of periodic, zero-mean functions with Sobolev regularity $2p$. 
It can be shown that this indeed defines a proper Gaussian prior $\Pi$ on $L^2[0,1]$. 
The hyper parameter $p$ determines the regularity of the prior in some sense. 
As shown in \cite{PokernStuartvanZanten}, the prior gives mass $1$ to a space of functions that 
have H\"older regularity $\alpha$ for every $\alpha < p-1/2$.

It turns out that for this prior we can explicitly derive the corresponding posterior. It is a Gaussian measure on $L^2[0,1]$
again and we can obtain expressions for posterior  mean and precision  operator. 
These expressions involve the periodic local time of the diffusion. This is the random field 
$L^\circ = (L^\circ_T(x): T \ge 0, x \in [0,1])$ with the property that 
for every $1$-periodic, nonnegative measurable function $f$ and $T > 0$, 
\[
\int_0^T f(X_t)\,dt = \int_0^1 f(x)L^\circ_T(x)\,dx.
\]
Moreover, we need the random field $\chi^\circ$ defined by 
\[
\chi^\circ_T(x) = 
\begin{cases}
\#\{k \in \ZZ: X_0 < x+k < X_T\} & \text{if $X_0 < X_T$},\\
-\#\{k \in \ZZ: X_T < x+k < X_0\} & \text{if $X_T < X_0$},\\
0 & \text{else}.
\end{cases}
\]

A non-rigorous derivation of the posterior now proceeds by first noting that 
by It\^o's formula and periodicity, the likelihood \eqref{eq: lik} can be written as 
\[
p(X\given b) = \exp\Big(-\frac{1}{2}
\int_0^1 \left( L_T^\circ(x) \left(b^2(x) +b'(x)\right)-
2 \chi_T^\circ(x)b(x) \right) dx\Big).
\]
Next we observe that, 
very loosely speaking,  the Gaussian prior $\Pi$ has a ``density'' 
proportional to 
\begin{equation*}
b \mapsto \exp\Big(-\frac12 {\int_0^1 b(x) \Lambda^{-1}b(x)\,dx}\Big).
\end{equation*}
But then  the posterior has a ``density'' that is proportional to the product of these two quantities. 
Using also integration by parts we see that this equals 
\[
\exp\Big(-\frac12 {\int_0^1 b(x) (\Lambda^{-1}+ L_T^\circ(x))b(x)\,dx}
+\int_0^1 b(x)\Big(\frac{1}{2} (L^\circ_T(x))' + \chi_T^\circ(x)\Big)\,dx\Big).
\]
This is, up to a constant, again the ``density'' of a Gaussian measure. By 
completing the square we see that its mean $\hat b_T$ and covariance operator $\Lambda_T$ satisfy
$ \Lambda_T^{-1} = \Lambda^{-1}+L^\circ_TI$ and 
$ \Lambda_T^{-1} \hat b_T = \frac{1}{2} (L^\circ_T)' + \chi_T^\circ$. 
Recalling the definition of $\Lambda$ we obtain the relations
\begin{align*}
 \Lambda_T^{-1} &= \eta\left(-\Delta\right)^p + \left(\eta\delta +L^\circ_T\right)I,\\
 \eta\left(-\Delta\right)^p\hat b_T + (\eta\delta+L^\circ_T)\hat b_T &= \frac{1}{2} (L^\circ_T)' + \chi_T^\circ. 
\end{align*}

To make the derivation of the posterior rigorous the above differential equation for the posterior mean 
has to be understood in an appropriate weak sense, since the ordinary derivative of local time does not exist.
As detailed in \cite{PokernStuartvanZanten} this is indeed possible and it can be shown that the 
differential equation for $\hat b_T$ has a unique weak solution. Well-established 
methods from numerical analysis can be then used to compute the posterior. For further details 
and application of the approach in problems from molecular dynamics and finance, we
refer to \cite{Papa}.

The approach of \cite{Papa} can be compared to the naive approach outlined in the preceding subsection 
by noting that their prior $\Pi$ can in fact equivalently be described by a series expansion. 
Define the basisfunctions
\begin{align*}
\psi_{2k}(x) & =\sqrt{2} \cos(2\pi kx),\\
\psi_{2k-1}(x) &=\sqrt{2} \sin(2\pi kx), 
\end{align*}
 for $k \in \mathbb{N}$. It is easily verified that the functions 
 $\psi_k$  are the eigenfunctions of  the prior precision operator $\Lambda^{-1}$, and the corresponding eigenvalues 
 are given by 
\begin{equation*}
\lambda_k^{-1} = \eta \left( 4\pi^2 \left\lceil \frac{k}{2} \right\rceil^2 \right)^{p}
+\delta. 
\end{equation*}
It follows that the prior on $b$ can also be defined structurally by expanding $b = \sum_{k=1}^\infty c_k\psi_k$ and 
putting independent Gaussian priors on the coefficients $c_k$, with mean $0$ and variance $\lambda_k$.  

So compared to the naive finite series approach, the proposal of \cite{Papa} is genuinely nonparametric 
and  eliminates  the chance of missspecifying the form of the drift. 
On the down side, the prior has fixed hyper parameters that still might combine poorly with the true drift.
 In particular, there remains a possibility that a
multiplicative scaling parameter $\eta$ is chosen that incorrectly reflects the scale of the actual drift. 
This can deteriorate the quality of the inference. In the next subsection we discuss a prior which 
allows for a data-driven choice of the scaling.

\subsection{Infinite series priors}
\label{sec: moritz}

In this section we consider the approach proposed by \cite{Moritz}.
We consider the same setup as before, i.e.\ the drift $b$ is assumed to be $1$-periodic.

The prior of \cite{Moritz} is defined by  writing a truncated series expansion for $b$
and putting prior weights on the truncation point and on the coefficients in the expansion. 
We employ general
$1$-periodic, continuous basis functions $\psi_k$, $k \in \NN$. Next we fix an increasing sequence 
of natural numbers $m_j$, $j \in \NN$, to group the basis functions into {\em levels}. The 
functions $\psi_1, \ldots, \psi_{m_1}$ constitute level $1$, the functions $\psi_{m_1+1}, \ldots, \psi_{m_2}$ correspond to 
level $2$, etcetera. In this manner we can accommodate both families of basis functions with a single index (e.g.\ the Fourier basis)
and doubly indexed families (e.g.\ wavelet-type bases). Functions that are linear combinations 
of the first $m_j$ basis functions $\psi_1, \ldots, \psi_{m_j}$ are said to belong to {\em model} $j$. Model $j$ encompasses levels $1$ up till $j$. 

To define the prior on $b$ we first put a prior on the model index $j$, given by certain prior weights $p(j)$, $j \in \NN$. 
By construction, a function in model $j$ can be expanded as 
$\sum_{l=1}^{m_j} \theta^j_l\psi_l$
for a certain vector of coefficients $\theta^j \in \RR^{m_j}$. Given $j$, we endow this vector with a prior by postulating that  
the coefficients $\theta^j_l$ are given by an inverse gamma scaling constant times independent, centered Gaussians 
with certain decreasing variances. 

Concretely, to define the prior we fix model probabilities $p(j)$, $j \in \NN$, decreasing variances $\xi^2_l$, $l \in \NN$,
positive constants $a, b > 0$ and set $\Xi_j = \diag({\xi^2_1, \ldots, \xi^2_{m_j}})$.
Then the hierarchical prior $\Pi$ on the drift function $b$ is defined as follows:
\begin{center}
\parbox{0.6\textwidth}{
\begin{align*}
j & \sim p(j),\\
s^2 & \sim {\rm IG}(a,b),\\
\theta^j\given j, s^2 & \sim N_{m_j}(0, s^2\Xi_j),\\
b \given j, s^2, \theta^j & \sim \sum_{l=1}^{m_j} \theta^j_l\psi_l.
\end{align*}}
\end{center}
In the paper \cite{Moritz} particular choices for the basis functions, the prior on $j$ and the variances $\xi_l$ 
are considered in more detail.

This  prior is different from the one considered in the preceding section in a number of ways.
Firstly,  different basis functions can be used.
This added flexibility can be computationally attractive. 
The posterior computations involve the inversion of certain large matrices and choosing basis functions with 
local support  typically makes these matrices sparse.
A second difference is that we truncate the infinite series at a level that we endow with a prior as well. 
In this manner we can achieve considerable computational gains if the data driven truncation point is relatively small, 
so that only low-dimensional models are used and hence only relatively small matrices have to be inverted. 
A last important change is that we do not set the multiplicative hyper parameter at a fixed value, but instead
endow it with a prior and let the data determine the appropriate value. 

The simulations presented in  \cite{Moritz} indicate that 
this approach has several 
advantages. Although the truncation of the series at a data driven point 
involves incorporating reversible jump MCMC steps in the computational algorithm,  
it can indeed lead to a considerably faster procedure compared to truncating 
at some fixed high level. Moreover, the introduction of a prior on the multiplicative hyper parameter reduces 
the risk of misspecifying the scale of the drift. Using a fixed scaling parameter
can seriously deteriorate the quality of the inference, whereas the hierarchical procedure with a prior on that parameter
is able  to adapt to the true scale of the drift. 
Also, numerical investigations  indicate that by the introduction of a prior on the scale we also 
 can achieve some degree of adaptation to smoothness.

The prior $\Pi$ described above is constructed in such a way that numerical computation is practically feasible. 
Within a fixed model $j$, a Gibbs sampler for sampling $\theta^j$ and $s^2$ can be constructed using 
standard inverse gamma-normal computations.  
Reversible jump MCMC can be used to jump between different models. This involves the computation of 
certain Bayes factors, for which a closed form expression can be derived in this setup. If only low-frequency data
are available, the Gibbs and reversible MCMC steps can be combined with data augmentation steps involving 
the simulation of diffusion bridges, as outlined 
also in the preceding subsection.

In \cite{Moritz} the approach described above is applied to analyze the same butane dihedral angle time series as in \cite{Papa}. 
 After preliminary operations  these data can be considered to be  $4000$ observations from a 
 scalar diffusion with unit-diffusion coefficient, 
distributed evenly over the time  interval $[0,4]$.

\begin{figure}[h]
\begin{center}
Figure \ref{fig:pokern} about here.
\end{center}
\end{figure}

\begin{figure}[h]
\begin{center}
\includegraphics[width=.45\linewidth]{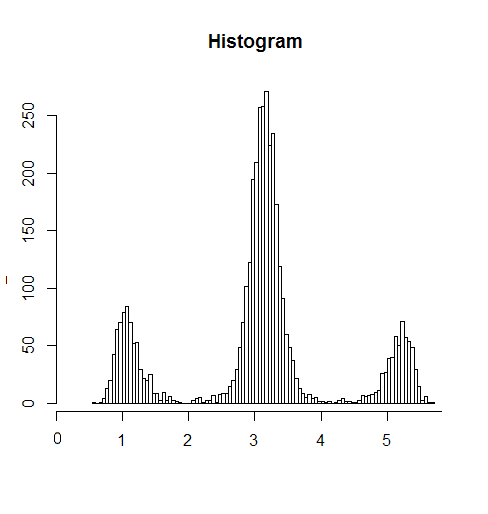}
\includegraphics[width=.45\linewidth]{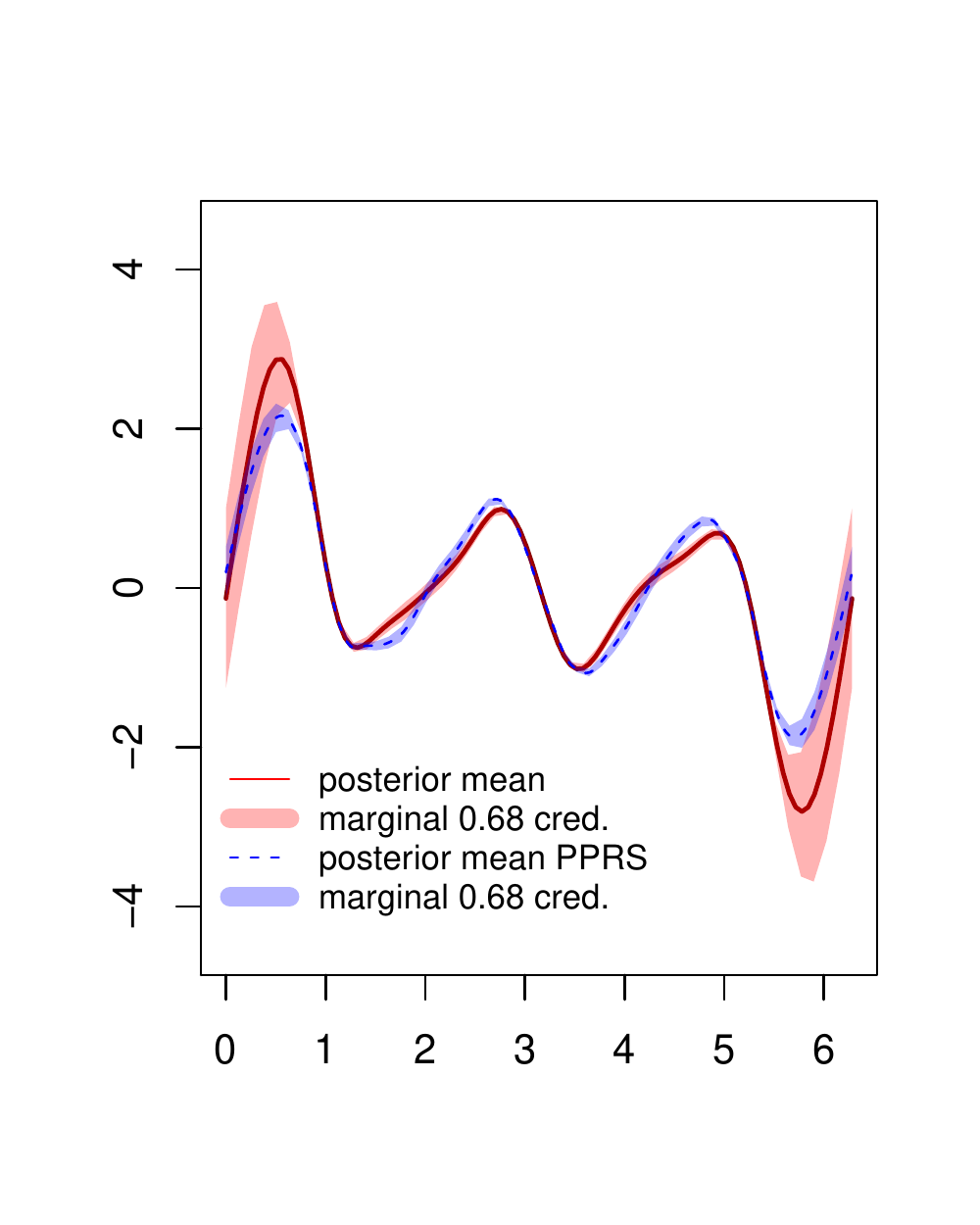}
\caption{Comparision of the estimate of drift using the Butane Dihedral Angle data. Left: Histogram of the data. 
Right:  posterior corresponding to prior from Section \ref{sec: moritz} in red and posterior   corresponding 
to prior from Section \ref{sec: papa} in blue.}\label{fig:pokern}
\end{center}
\end{figure}

The left panel of Figure \ref{fig:pokern} shows a histogram of the diffusion data. The right-hand panel shows the posterior mean 
and $68\%$ pointwise credible bands for the posterior distributions of the drift function corresponding to two different priors. The blue posterior 
corresponds to the prior of \cite{Papa} described
in Section \ref{sec: papa}, where we took, as in the cited paper, fixed hyperparameters $\eta=0.02$, $\kappa=0$ and $p=2$ in \eqref{eq:priorCov}. The red posterior corresponds to the prior described above, where we took Fourier basis functions and 
selected the hyperparameter to match the prior of \cite{Papa} as closely as possible. 
The fact that the red credible bands are wider near the boundary of the observation area, where less data are available,  indicates that 
\cite{Papa} might be somewhat overconfident about the form of the drift function in that area. 
Their narrower credible bands  seem to be  caused by prior belief rather than by information in the data and 
are not corroborated by the more conservative approach described in this section, which allows for a data-driven scaling 
constant and truncation point. 

For more details about  computational issues and simulation examples we 
refer to \cite{Moritz}.

\section{Asymptotics}
\label{sec: asymp}

The negative examples of e.g.\ \cite{Free63, Free65} or \cite{Diab, Diaa} show that in Bayesian nonparametrics,
even intuitively reasonable priors may lead to inconsistent procedures. More generally, it is by now well known that contrary 
to the parametric setting, the choice of the prior has a large impact on the performance in infinite-dimensional models. 
This performance is determined by  fine mathematical properties  and can not be assessed by simply
eyeballing the prior. 
As a result there is an interest in 
mathematical results that relate properties of the prior to the quality of the Bayes procedure. Such results 
can serve as guidelines for the selection or construction of priors. 

Mathematical results in this setting typically assume that the data are generated using  a true drift function $b_0$
and study if and how the posterior concentrates around $b_0$ as  more and more data become available. In the 
continuous-time case in which we observe the diffusion on a time interval $[0,T]$ this means we let $T \to \infty$. 
In the low-frequency case it simply means we let the number $n$ of observations tend to infinity. 
Posterior consistency is the property that the posterior indeed contracts around $b_0$, in the sense that 
asymptotically, any neighborhood (relative to a suitably define topology) of $b_0$ receives posterior mass $1$. 
This is a  property that any reasonable procedure should ideally have.  Once posterior consistency has been established, 
the rate at which the posterior contracts around the true $b_0$ can be studied. In particular it can be investigated 
whether a certain prior leads to optimal convergence rates. 

For  diffusion models, 
the  paper \cite{MeulenVaartZanten} was the first to systematically study convergence rates
for nonparametric Bayes procedures. 
General conditions were derived for attaining a certain rate of contraction, in terms of the metric entropy of 
the support of the prior and the mass that the prior assigns to neighborhoods of the true function. These 
conditions are the analogues of similar conditions that were initially derived for the setting of i.i.d.\ density estimation 
by \cite{GGvdV}.  A concrete prior for ergodic diffusions considered in \cite{MeulenVaartZanten}
is a Dirichlet process like prior designed to model decreasing drift functions. This prior is shown to attain the 
optimal convergence rate $T^{-1/3}$ (up to a logarithmic factor).
Certain Gaussian process priors for the drift, essentially multiply integrated Brownian motions, are 
considered in the paper \cite{PanzarVZanten}. These priors have a certain fixed degree of regularity. Brownian 
motion has smoothness of order $1/2$, integrated Brownian motion has smoothness level $3/2$, etc. 
For such Gaussian priors the message is that if the regularity of 
the prior that is used coincides with the regularity of the unknown drift function, then optimal contraction rates are achieved.
Specifically, if both the drift and the prior have regularity $\beta > 0$, then the posterior contracts 
around the true drift at the rate $T^{-\beta/(1+2\beta)}$, which can be shown to 
be optimal in a certain sense, cf.\ e.g.\ \cite{Kut}. 
If the two regularities are not equal however, only sub-optimal speeds are realized in general. 
This is in line with general findings for Gaussian priors in other settings, see e.g.\ 
\cite{vdVvZ}, \cite{Ismael}. 

The concrete Gaussian prior with precision operator \eqref{eq:priorCov} described in Section \ref{sec: papa} is analyzed in detail in the paper \cite{PokernStuartvanZanten}. As mentioned above, this prior has regularity $\beta = p-1/2$. 
It is proved in \cite{PokernStuartvanZanten} that the corresponding posterior contracts around the true drift at the rate 
$T^{-(2p-1)/(4p)}$, provided the drift has regularity $p$. Note that this rate is also the optimal $T^{-\beta/(1+2\beta)}$, 
but the assumption on the drift is that it is $\beta+1/2$-regular. 
It is expected however that also in the periodic setting of \cite{PokernStuartvanZanten} this assumption on the drift can be 
weakened to $\beta$-regularity, and that just as in the ergodic setting studied in \cite{PanzarVZanten}, 
it holds that a Gaussian prior with fixed regularity is rate-optimal if and only if its smoothness matches the smoothness 
of the true drift. 

Priors that perform optimally across a whole range of regularities for the drift, i.e.\ so-called rate-adaptive priors 
have not yet been exhibited for diffusion models. It is expected that the prior of \cite{Moritz} described in Section 
\ref{sec: moritz} allows for a degree of adaptation to smoothness, but this has not yet been established. 
A combination of the general theory of \cite{MeulenVaartZanten} and new  local time asymptotics obtained in \cite{PokernStuartvanZanten} 
are expected to shed further light on the matter in the future. 

The asymptotic results mentioned thus far all concern continuously observed diffusions, where the accumulation 
of information is ensured either by ergodicity or by periodicity assumptions. 
The derivation of usable results for the low-frequency setting is much more involved. The fact 
that the discrete-time likelihood can not be explicitly expressed in terms of the drift complicates the analysis
considerably. At the present time, the only available results deal with posterior consistency  relative to 
a weak topology, cf.\  \cite{VdMeulenVZanten} and the extensions in \cite{Shota}. 
It is a great challenge to obtain consistency results for stronger topologies and rate of contraction results 
for procedures based on low-frequency data.

\section{Concluding remarks}
\label{sec: con}

Nonparametric Bayesian methodology for stochastic differential equations has started to 
develop only very recently. 
At this point in time there exist only a few methods that are computationally feasible. Moreover, the 
theoretical performance analysis of these methods is still rather immature. 
Nevertheless, the nonparametric Bayes approach can be expected to become more and more 
common in the near future, since it combines the advantages of flexible, nonparametric modeling with 
the possibility of providing uncertainty quantification and the possibility to deal with low-frequency data. 
This development will be stimulated by ongoing work on computational matters and theoretical foundations. 

For multi-dimensional diffusions, nonparametric Bayes methods have until now not been developed and 
studied. The challenge in higher dimensions are even larger than in the scalar case. In particular, 
it is not straightforward to generalize the diffusion bridge approach. In principle diffusion bridges
can be constructed in higher dimensions as well (e.g.\ \cite{DelyonHu}), but typically numerical implementations are 
far too slow to be useful in the nonparametric case.
  For parametric 
models it is already rather difficult to devise computationally feasible and theoretically sound methods  
in this case. It is not a-priori clear how approaches that have proven to be successful in the parametric 
setting (e.g.\ \cite{GW1}, \cite{GW2}) can be adapted to the nonparametric case.

Another challenge that remains is dealing with unknown parameters in the diffusion coefficient, in particular
in combination with generating missing sample paths between the data points (\cite{RobertsStramer}). Some promising 
ideas are being developed (e.g.\ \cite{StramerBognar}), but much work remains to be done.


\end{document}